\documentclass[12pt]{article}

\newcommand{\ampm}{A_{\mu}^{\pm}}
\newcommand{\ampmp}{A_{\mu}^{\pm\,\prime}}
\newcommand{\ampmpp}{A_{\mu}^{\pm\,\prime\prime}}
\newcommand{\amth}{A_{\mu}^{3}}
\newcommand{\amthp}{A_{\mu}^{3\,\prime}}
\newcommand{\amthpp}{A_{\mu}^{3\,\prime\prime}}
\newcommand{\be}{\begin{equation}}
\newcommand{\ee}{\end{equation}}

\newcommand{\lath}{\lambda_{3}}

\newcommand{\laei}{\lambda_{8}}
\newcommand{\sth}{$SU(3)\,$}
\newcommand{\stw}{$SU(2)\,$}

\usepackage{graphicx}

\begin{document}
\pagestyle{plain}
\begin{titlepage}

\begin{tabbing}
\` ILL-(TH)-01-10 \\
\` October  2001 \\
\` revised, May  2002 \\
\end{tabbing}
 
 
\begin{center}
{\bf The Maximal Abelian Gauge, Monopoles, and Vortices\\
in SU(3) Lattice Gauge Theory \\}

\vspace*{.5in}

John D. Stack and William W. Tucker\\
{\it Department of Physics \\
University of Illinois at Urbana-Champaign \\
1110 W. Green Street \\
Urbana, IL 61801 \\}

\vspace*{0.2in}

and

\vspace*{0.2in}
Roy J. Wensley\\
{\it Department of Physics and Astronomy\\
Saint Mary's College\\
Moraga, CA 94575 \\}
\end{center}

\begin{center}
{\bf Abstract}
\end{center}

\noindent

We report on calculations of the heavy quark potential in $SU(3)$ lattice
gauge theory.  Full $SU(3)$ results are compared to three cases which
involve gauge-fixing and projection.  All of these start from the maximal
abelian gauge (MAG), in its simplest form. The first case is 
abelian projection to $U(1) \times U(1)$.   
The second keeps only the abelian fields of  monopoles
in the MAG.  The third involves 
an additional 
gauge-fixing to the {\mbox indirect} maximal center gauge (IMCG),  
followed by center
projection to $Z(3)$.  At one gauge fixing/configuration, the
string tensions calculated from MAG $U(1)\times U(1)$, MAG monopoles,
and IMCG $Z(3)$ are all less than the full $SU(3)$ string tension.  
The  projected string tensions further decrease, by approximately
10\%, when account is taken of gauge ambiguities.  Comparison is made with
corresponding results for $SU(2)$.  It is emphasized that the formulation of
the MAG is more subtle for $SU(3)$ than for $SU(2)$, and that the low
string tensions  may be caused by the simple MAG form used.  A generalized
MAG for $SU(3)$  is formulated.

\begin{tabbing}
\` PACS Indices: 11.15.Ha\\
\` 12.38.Gc\\
\end{tabbing}

\end{titlepage}

\pagestyle{plain}
\section{Introduction}
One  of the most important ideas in 
the theory of quark confinement is the notion
of abelian projection, introduced long ago by `t Hooft \cite{thooft}.
The basic conjecture made by `t Hooft was that in an appropriate
partially fixed gauge, confinement can be understood for $SU(N)$
pure gauge theory by retaining only the gauge fields belonging to the
maximal abelian subgroup $U(1)^{N-1}$.  
In `t Hooft's original paper, 
there were several suggestions for how to partially
fix the gauge, leaving the maximal abelian subgroup unfixed.  Subsequent
work in $SU(2)$ lattice gauge theory has focussed on
the well known `maximal abelian gauge' (MAG).  The main purpose of the
present work is to perform  lattice gauge theory 
calculations which test `t Hooft's idea of abelian projection for $SU(3)$,
using the simplest possible form of the MAG.

\subsection{Beyond Abelian Projection}
The projection process can be carried beyond abelian projection,
keeping a still smaller
set of degrees of freedom.  Closely related to abelian projection or
abelian dominance is the postulate of monopole dominance.  This assumes
that the confining physics in the abelian gauge fields originates in the
magnetic current of monopoles. The monopoles are assumed to be 
similar in nature to 't Hooft-Polyakov
monopoles, moving on world lines.  It is known that a dense network
of magnetic current is the agent of confinement in 
$U(1)$ lattice gauge theory \cite{js-rw}.  
The key question is whether  a corresponding network
can explain confinement in $SU(N)$ lattice gauge theory.

Another possibility is that the confining degrees of freedom are really
associated with the $Z(N)$ center of $SU(N)$.  Since 
$Z(N) \subset U(1)^{N-1}$, this can be studied 
{\mbox  after} making
an abelian projection to $U(1)^{N-1}$.  
After the initial gauge-fixing and abelian projection,   
the $U(1)^{N-1}$ abelian links are subjected to a second gauge-fixing and
projection, leaving a set of $Z(N)$ or center-valued links.  This is
the `indirect maximal center gauge' (IMCG). The IMCG uses the MAG as an
intermediate stage, as opposed to the `direct maximal center gauge' (DMCG),
which does not.  For either case, center dominance is tested by using
$Z(N)$ links, or closed world sheets of plaquettes carrying 
$Z(N)$ flux.  The latter are called (P)-vortices.

\subsection{ Gauge Ambiguities and  SU(2) vs. SU(3)}
In this paper, we work exclusively 
with `maximal' gauges;  gauges in which
a certain functional of the links is maximized before the links are
abelian or center-projected.  The use of a maximal gauge has an intuitive
appeal.  In the case of abelian projection, if the charged degrees of
freedom are to be deleted, why not first transform to a gauge where they
are made as small as possible? Maximal gauges also have the desirable practical feature that the noise level in projected
configurations  is quite low.  

A complication associated with maximal gauges is the presence of
gauge or Gribov ambiguities.  The functionals used have many local maxima
for a given configuration.
Each local maximum of the gauge functional is a `gauge copy',  and the
results of numerical calculations depend on which gauge copy is chosen
for each configuration.
That this variation with gauge copies occurs at all
is caused by projection.
If the complete set of gauge degrees of freedom is kept, 
a gauge invariant quantity like  a
Wilson loop has the same value  at any point on the 
gauge orbit of a configuration.  But if 
certain degrees of freedom are deleted, then the  value of  the
projected Wilson loop will in general be different at different points on
the orbit, even if the chosen points are 
 local maxima of a gauge functional.  

\stw calculations have shown a  clear qualitative trend for this variation
due to gauge copies.  Gauge copies with higher values of the
functional produce projected configurations which are more ordered 
and produce
lower string tensions.
This is true for  
the maximal abelian gauge followed by abelian projection,  as well as
the direct maximal center gauge 
followed by center projection. In both cases,  if for each configuration  a
randomly chosen gauge copy is used, the projected string tension turns out to be
larger than the full $SU(2)$ result.  If instead, a collection of gauge copies
is generated for each configuration, selecting a subset and then picking the
member of the subset with the highest value of the gauge functional leads
to a reduced string tension.
For the MAG and abelian projection, recent work  suggests that use of the
`best' (global maximum) gauge copy for each 
configuration would lead to an 
abelian string tension quite close to the full 
$SU(2)$ value \cite{bornyakovsu2}.
For the DMCG and center projection, there is more variation.
As gauge copies with increasing values of the center gauge functional are
used, the $Z(2)$ string tension decreases from an initial  value 
above the full $SU(2)$ string tension to a value
well below it \cite{bornyakov}.
While the global maximum of the gauge functional 
evidently plays a different role in the two
projection schemes,
for the purposes of this paper, the important point is that for both
abelian and center projection, there exist gauge copies 
that give string tensions very close to the exact $SU(2)$ value.

For $SU(3)$,  
the correct formulation of the MAG condition is a more subtle matter 
than it is for \stw.  This is discussed in detail in Sec.(\ref{sec_mag}).
The fact that there is more than one possible form for the MAG 
is important, since one of
our main results is that for the simplest form of the MAG in \sth, it is
practically certain that no gauge 
copies exist which produce a $U(1)\times U(1)$
string tension close to the full \sth value.  

Maximal center gauge
conditions are discussed in Sec.(\ref{sec_center}).  In our numerical work, we
used the IMCG
which has the MAG as a preliminary step.
Here we also find a low string tension.  The amount by which the $Z(3)$
string tension is low depends on the treatment of gauge copies.
Since
the MAG and abelian projection are an intermediate stage, the
low values we find for the $Z(3)$ string tension are not necessarily 
a problem of center projection in $SU(3)$, but may be reflecting 
the difficulty discussed above for the simple form of the MAG.

Our numerical calculations were performed  at $\beta=5.9$
on a $10^{3} \times 16$ lattice and at
$\beta=6.0$ on a $16^4$ lattice.
The full \sth results are in Sec.(\ref{full_su3}).  For the  MAG, the
$U(1)\times U(1)$ results are in Sec.(\ref{sec_u1u1}),
and monopole results are in Sec.(\ref{sec_monoresults}).  For the IMCG, the
$Z(3)$ results are in Sec.(\ref{sec_z3results}).  We find that as in
$SU(2)$, the monopole and center vortex degrees of freedom are intimately
connected.  At $\beta=6.0$,  for both the MAG and IMCG,
we investigated the dependence on gauge copies.  We confirm for both gauges,
that gauge copies with higher values of the gauge functional 
lead to lower string tensions.  This is the same general 
behavior as seen in $SU(2)$.
Sec.(\ref{sec_conc_js}) contains
a summary and conclusions.

\pagestyle{plain}
\section{Maximal Abelian Gauge}
\label{sec_mag}
In this section, we discuss the maximal abelian gauge condition for $SU(3)$,
comparing to the familiar case of $SU(2)$.  For $SU(2)$, we regard 
$A_{\mu}^{3}$ as the abelian field, and by making gauge transformations,
suppress the `charged' fields
$A_{\mu}^{1}$ and $A_{\mu}^{2}$  to the greatest extent possible.  The 
continuum functional
\be
G_{mag}^{su2}=\int d^{4}x[(A_{\mu}^{1})^{2}+(A_{\mu}^{2})^{2}]
\label{su2mag}
\ee
is minimized over $SU(2)$ gauge transformations.  
Near a minimum, $G_{mag}^{su2}$ is stationary 
when the gauge fields are varied by
an infinitesimal gauge transformation,
\be
A_{\mu}^{a}  \rightarrow A_{\mu}^{a} - \partial_{\mu}\theta^{a}
+\epsilon^{abc}\theta^{b}A_{\mu}^{c}.
\ee
Since $G_{mag}^{su2}$ is 
invariant to a small abelian gauge transformation
involving $\theta^{3}$, only transformations involving $\theta^{1},
\theta^{2}$ lead to gauge conditions. 
If we combine the gauge fields  into charged 
combinations, $\ampm =(A_{\mu}^{1} \pm iA_{\mu}^{2})/\sqrt{2}$, 
the conditions  can be written as
\be
\partial_{\mu}\ampm \pm i\amth \ampm =0.
\label{su2mageq}
\ee

Turning to $SU(3)$, there are now two abelian gauge fields, $A_{\mu}^{3}$ and
$A_{\mu}^{8}$.  The remaining six gauge fields are all charged with respect
to at least one of $A_{\mu}^{3},A_{\mu}^{8}$.  Treating these charged fields
democratically, the simplest maximal abelian gauge condition for $SU(3)$ would
minimize the functional
\be
G_{mag}^{su3}=\int d^{4}x[(A_{\mu}^{1})^{2}+(A_{\mu}^{2})^{2}
+(A_{\mu}^{4})^{2}+(A_{\mu}^{5})^{2}+
(A_{\mu}^{6})^{2}+(A_{\mu}^{7})^{2}]
\label{su3magsimp}
\ee
over  the $SU(3)$ gauge group.  An infinitesimal $SU(3)$ gauge
transformation is
\be
A_{\mu}^{a}  \rightarrow A_{\mu}^{a} - \partial_{\mu}\theta^{a}
+f^{abc}\theta^{b}A_{\mu}^{c},
\ee
where $f^{abc}$ are the \sth structure constants.  Since
$G_{mag}^{su3}$
is invariant to a small abelian gauge transformation involving
$\theta^{3}$ and $\theta^{8}$, there are six gauge conditions 
arising from the remaining six angles. 
These conditions can  easily be expressed in terms of the three
$SU(2)$ subgroups of \sth.  
We denote the three subgroups and their gauge fields as follows:
\begin{eqnarray*}
I:          & A_{\mu}^{1},A_{\mu}^{2},\amth & \\
V: & A_{\mu}^{4},A_{\mu}^{5},\amthp & \\
U: & A_{\mu}^{6},A_{\mu}^{7},\amthpp, & 
\end{eqnarray*}
where $\amthp=\frac{1}{2}A_{\mu}^{3}+\frac{\sqrt{3}}{2}A_{\mu}^{8}$,
$\amthpp=-\frac{1}{2}A_{\mu}^{3}+\frac{\sqrt{3}}{2}A_{\mu}^{8}$, and
with some abuse of notation,
quark model terminology has been used to denote the subgroups.
Consider the response of $G_{mag}^{su3}$ to an infinitesimal 
gauge transformation in the $I$-spin subgroup.  Only the terms
in  $G_{mag}^{su3}$ involving $A_{\mu}^{1},A_{\mu}^{2}$ vary under this
transformation, so  demanding that $G_{mag}^{su3}$ be stationary leads to
Eq.(\ref{su2mageq}) again. Likewise, demanding that
$G_{mag}^{su3}$ be stationary with respect to a $V$-spin  gauge
transformation leads to 
\be 
\partial_{\mu}\ampmp \pm i \amthp\ampmp=0,
\label{su2mageqp}
\ee
where $\ampmp=(A_{\mu}^{4} \pm i A_{\mu}^{5})/\sqrt{2}$.
Finally,  $U$-spin gauge transformations lead to
\be 
\partial_{\mu}\ampmpp \pm i \amthpp\ampmpp=0,
\label{su2mageqpp}
\ee
where $\ampmpp=(A_{\mu}^{6} \pm i A_{\mu}^{7})/\sqrt{2}$.
So the $SU(3)$ maximal abelian gauge functional Eq.(\ref{su3magsimp})
leads to  the requirement
that the gauge fields be `maximally abelian' with respect to each of the three
$SU(2)$ subgroups of $SU(3)$.

Minimizing the continuum functional
Eq.(\ref{su3magsimp}) is equivalent to maximizing
the lattice functional ${\bf G}_{mag}^{su3}$ defined by
\begin{eqnarray}
{\bf G}_{mag}^{su3}
&=&\sum_{x,\mu}\left({\rm tr}(U_{\mu}(x)\lambda_{3}\right.
U_{\mu}^{\dagger}(x)\lambda_{3})+
\left.{\rm tr}(U_{\mu}(x)\lambda_{8}U_{\mu}^{\dagger}(x)\lambda_{8})\right)
\label{lattmag}\\
&=&
2\sum_{x,\mu}
\left(\left|(U_{\mu})_{11}\right|^{2}+\left|(U_{\mu})_{22}\right|^{2}\right.
\left.+\left|(U_{\mu})_{33}\right|^{2}-1\right) \nonumber.
\end{eqnarray}
We will use Eq.(\ref{lattmag}) for our MAG gauge-fixing in this paper. 
This same functional has been used in the small number of other papers
published on the MAG for $SU(3)$ \cite{schierholz1,schierholz2}.

At a local maximum of  Eq.(\ref{lattmag}), three conditions are satisfied.
These are the lattice analogs of  the 
continuum conditions 
Eqs.(\ref{su2mageq}),(\ref{su2mageqp}), and (\ref{su2mageqpp}), 
and are satisfied when
${\bf G}_{mag}^{su3}$ is stationary with respect to
 gauge transformations  in the
$I,U$, and $V$-spin subgroups. To write the conditions explicitly,
we define three site-valued matrices.  For the $I$-spin
subgroup, the matrix ${\bf X}(x)$ is
\be
{\bf X}(x)\equiv \sum_{\mu}\left[U_{\mu}(x)\lambda_{3}U^{\dagger}_{\mu}(x)
+U^{\dagger}_{\mu}(x-\hat{\mu})\lambda_{3}U_{\mu}(x-\hat{\mu})\right]
\label{gcond}.
\ee
For the $V$ and $U$-spin subgroups, the
corresponding matrices ${\bf X}^{\prime}$ and ${\bf X}^{\prime\prime}$
are obtained by replacing $\lambda_{3}$ in Eq.(\ref{gcond}) by
$\lambda^{\prime}_{3}$ and $\lambda^{\prime\prime}_{3}$, respectively,
where
\be
\lambda_{3}^{\prime} \equiv
\frac{1}{2}\lambda_{3}+\frac{\sqrt{3}}{2}\lambda_{8},\;
\lambda_{3}^{\prime\prime} \equiv
-\frac{1}{2}\lambda_{3}+\frac{\sqrt{3}}{2}\lambda_{8}.
\ee
When ${\bf G}_{mag}^{su3}$ is stationary against gauge  transformations
in the $I$-spin subgroup, the 12 and 21 matrix
elements of ${\bf X}(x)$ must vanish; similarly, 
for the $V$-spin subgroup, the
13 and 31 matrix elements of ${\bf X}^{\prime}(x)$ must vanish; and for
the $U$-spin subgroup, the 23 and 32 matrix elements of
${\bf X}^{\prime\prime}(x)$ must vanish.  In practice, we require that
these matrix elements be very small.
The numerical methods we use to reach these conditions 
with high accuracy will be discussed in Sec(\ref{sec_u1u1}).

\subsection{Generalized Maximal Abelian Gauge}
\label{mag_gen}
So far there appears to be a close analogy between the MAG conditions for
$SU(2)$ and $SU(3)$.  However, there are more general possibilities for
$SU(3)$.  To see them, we first note that $G_{mag}^{su2}$  is in the form
of a mass term for the gauge fields $A_{\mu}^{1},A_{\mu}^{2}$.
Such a term would arise in a gauge-Higgs theory, 
with an adjoint
Higgs field $\Phi^{a}$  coupled 
to the $SU(2)$ gauge field.  
If the Higgs field has a vacuum expected value along the
3-axis, equal mass terms for $A_{\mu}^{1}$ and $A_{\mu}^{2}$ will be generated.
Take the nondynamical Higgs field to have a 
fixed length, $\Phi^{a}\Phi^{a}=\Phi_{0}^{2}$, and write the continuum
Higgs action
\be
G_{higgs}^{su2}=
\frac{1}{2}\int d^{4}x D_{\mu} \vec{\Phi} \cdot D_{\mu} \vec{\Phi},
\label{su2higgs}
\ee
where 
$(D_{\mu}\Phi)^{a}= \partial_{\mu}\Phi^{a}+ \epsilon^{abc}A_{\mu}^{b}\Phi^{c}$.
In the Higgs gauge with  $\vec{\Phi}=\Phi_{0}\hat{3}$, $G_{higgs}^{su2}$
becomes
\be
G_{higgs}^{su2}=\frac{\Phi_{0}^{2}}{2}\int\left( (A_{\mu}^{1})^{2} \right.
\left. +(A_{\mu}^{2})^{2} \right) d^{4}x
\ee
which is up to a constant, the same as $G_{mag}^{su2}$.

The \sth functional Eq.(\ref{su3magsimp}) is also in the form of  mass
terms for the charged gauge fields.
It is  of interest to see how the functional
Eq.(\ref{su3magsimp}) is related to the action of a fixed length adjoint
Higgs field $\vec{\Phi}$, with components $\Phi^{a}$.  We write the 
continuum action
\be
G_{higgs}^{su3}=
\frac{1}{2}\int d^{4}x D_{\mu} \vec{\Phi} \cdot D_{\mu} \vec{\Phi},
\label{su3higgs}
\ee
where
$(D_{\mu}\Phi)^{a}= \partial_{\mu}\Phi^{a}+ f^{abc}A_{\mu}^{b}\Phi^{c}$.
For $SU(3)$, going to a Higgs gauge means putting the Higgs field in
the Cartan algebra, 
$\vec{\Phi}=\Phi_{0}(\hat{3}\cos\chi+\hat{8}\sin\chi)$, 
or in matrix form 
\mbox{$\Phi=\Phi_{0}(\lath \cos\chi+\laei \sin\chi)/2$.}
The value of the angle 
$\chi$ cannot be changed by an \sth gauge transformation,
so there are distinct `moduli' or Higgs gauges parameterized by $\chi$. 
This is a difference between \stw and \sth.
In a generalized Higgs 
gauge where we allow $\chi$ to depend on $x$,
 the functional Eq.(\ref{su3higgs}) takes the form
\begin{eqnarray}
G_{higgs}^{su3} & = & \frac{1}{2}\Phi_{0}^{2}\int d^{4}x
\left\{ \partial_{\mu}\chi \partial_{\mu} \chi+ \right.
\cos^{2}(\chi)\left[(A_{\mu}^{1})^{2}+(A_{\mu}^{2})^{2}\right]
+\nonumber\\
 &  &
\cos^{2}(\frac{\pi}{3}-\chi)\left[(A_{\mu}^{4})^{2}+(A_{\mu}^{5})^{2}\right]
+\cos^{2}
(\frac{\pi}{3}+\chi)\left[(A_{\mu}^{6})^{2}\right.
\left.\left.+(A_{\mu}^{7})^{2}\right]\right\}.
\label{su3maghiggs}
\end{eqnarray}
Even for $\chi$ independent of $x$, this 
is clearly different from Eq.(\ref{su3magsimp}).  No individual adjoint
Higgs field can give equal coefficients for the 
various charged vector fields
in the functional. 
Formally, equal coefficients for the
gauge fields can be obtained by taking $\chi$ to be independent of $x$,
and then averaging over $\chi$  in Eq.(\ref{su3maghiggs}).

Minimizing the functional Eq.(\ref{su3maghiggs}) in the continuum
is equivalent on the lattice to maximizing  the functional 
${\bf G}_{higgs}^{su3}$
defined by
\be
{\bf G}_{higgs}^{su3}
=\sum_{x,\mu} {\rm tr}((\lambda_{3}\cos\chi+\lambda_{8}\sin\chi)U_{\mu}(x)
\left.(\lambda_{3}\cos\chi^{\prime}+\lambda_{8}\sin\chi^{\prime})U_{\mu}^{\dagger}(x)\right),
\label{latthiggs}
\ee
where $\chi=\chi(x)$ and $\chi^{\prime}=\chi(x+\hat{\mu}a)$. 
If $\chi$ is taken to be independent of $x$, and  ${\bf G}_{higgs}^{su3}$ is
averaged over $\chi$, ${\bf G}_{higgs}^{su3}$
reduces to ${\bf G}_{mag}^{su3}$.
Although not used in the present paper,
the functional ${\bf G}_{higgs}^{su3}$ would appear  to
be an interesting alternative to
${\bf G}_{mag}^{su3}$, particularly for the case where $\chi$ is allowed to 
depend on $x$.  We intend to explore this in future work.

\subsection{Abelian Projection}
\label{abel_proj}
After the MAG gauge-fixing process is complete on 
a given configuration, each link $U_{\mu}$ is factored into a diagonal
$U(1) \times U(1)$ part $u_{\mu}$, times a `charged' part $U_{\mu}^{ch}$, so
\mbox{$U_{\mu}=u_{\mu}U_{\mu}^{ch}$.}  The $u_{\mu}$ has generators in the 
Cartan algebra and represents the `phase' of
the full \sth link matrix $U_{\mu}$.  Short-range physics is hopefully
isolated in the  matrix $U_{\mu}^{ch}$, which
depends on the remaining six angles. In contrast to the \stw case,
simply extracting the phases of the diagonal elements of $U_{\mu}$
will not result in a
matrix $u_{\mu}$ with $\det(u_{\mu})=1$.  In this paper, we use two
different methods  to extract
$u_{\mu}$.  In the `symmetric' method, the phase
of a given element of $u_{\mu}$ is set equal to the phase of the corresponding
diagonal element of $U_{\mu}$, minus the average of the three 
diagonal phases.  In the second, `optimal' method, we find the $U(1) \times
U(1)$ link which maximizes ${\rm Re( tr(}u_{\mu}U_{\mu}^{\dagger})$
\cite{stacktucker3}.  
This finds the
`best' $U(1)\times U(1)$ approximation to $U_{\mu}$, and requires
an iterative procedure.  We have carefully checked that results produced by
symmetric or optimal methods differ by a negligible amount, considerably
smaller than statistical errors.

Once the $U(1)\times U(1)$ links $u_{\mu}$ are determined,  the 
abelian projection is carried out by setting
$U_{\mu}^{ch}$ to the identity on every link of the lattice, so
the resulting $SU(3)$ links are equal to the 
$u_{\mu}$.  Using the $u_{\mu}$ in place of the $U_{\mu}$ directly
tests abelian dominance for $SU(3)$.  The results of these calculations
are reported in Sec.(\ref{sec_u1u1}).  

\subsection{SU(3) Monopoles}
\label{mono_loc}

In this section, we discuss the extraction of
magnetic currents and calculations with monopoles for $SU(3)$, after reviewing
the  \stw case  \cite{stackwensleyneiman}.
There, after MAG gauge-fixing and abelian projection,
the links are diagonal $2 \times 2$ matrices 
$u_{\mu}$ given by
\be
u_{\mu}=
\left(
\begin{array}{cc}
\exp(i\phi_{\mu}^{11})  & 0   \\
0 & \exp(i\phi_{\mu}^{22})   \\
\end{array} \right ),
\label{su2diag}
\ee
where $\phi_{\mu}^{jj}\; \epsilon \; (-\pi,+\pi),\,j=1,2$ and 
 $\phi_{\mu}^{11}+\phi_{\mu}^{22}=0$.  Applying the Toussaint-DeGrand
procedure \cite{degrand} to plaquettes formed from the $\phi_{\mu}^{jj}$ gives 
two integer-valued magnetic currents $m_{\mu}^{j}$.
However,  $m_{\mu}^{1}+m_{\mu}^{2}=0$ is satisfied explicitly on every link, 
so there is only one
magnetic current for $SU(2)$.

For \sth , after going to the MAG and projecting to $U(1) \times U(1)$, the
abelian links are of the form
\be
u_{\mu}=
\left(
\begin{array}{ccc}
\exp(i\phi_{\mu}^{11})  & 0  & 0 \\
0 & \exp(i\phi_{\mu}^{22})  & 0 \\
0  & 0 & \exp(i\phi_{\mu}^{33}) 
\end{array} \right ),
\label{su3diag}
\ee
with $\phi_{\mu}^{11}+\phi_{\mu}^{22}+\phi_{\mu}^{33}=0,\,{\rm mod}(2\pi)$.  
Our procedure for extracting magnetic currents closely follows that 
just discussed for \stw.
Plaquette angles 
are computed for each color, and the Toussaint-DeGrand
procedure is applied  to each.
This produces a magnetic current for each color,
$m_{\mu}^{j},\, j=1,2,3$.  
Because of the ${\rm mod}(2\pi)$ in the sum of the angles $\phi_{\mu}^{jj}$,
the total magnetic current $m_{\mu}^{1}+m_{\mu}^{2}+m_{\mu}^{3}$ does not
vanish link by link.   However, as shown in Sec.(\ref{sec_monoresults}),
the total magnetic current has no physical effect, so there are actually only
two independent physical magnetic currents.
This method of treating the magnetic current has the advantage that it does
not force a particular subgroup structure for $SU(3)$ monopoles.  Evidence
that they are in fact associated with $SU(2)$ subgroups is presented
elsewhere \cite{jsstara}.

The monopole Wilson loops are 
calculated exactly as in \stw or $U(1)$ \cite{js-rw}.  
Each color
of magnetic current  produces a lattice magnetic vector potential
$A_{m\mu}^{j}$, and field strength 
$F_{m\mu\nu}^{j}=\partial_{\mu}^{+}A_{m\nu}^{j}
-\partial_{\nu}^{+}A_{m\mu}^{j}$.  The monopole Wilson loop is  then obtained
from the flux of  the dual of $F_{m\mu\nu}^{j}$ through the loop.  We have 

\begin{equation}
\left< W_{mon}^{j} \right>=
                \left<\exp\left(i\,2\pi\sum_{x,\mu > \nu}
D_{\mu\nu}(x)
{\raise0.20ex\hbox{*}} F_{m\mu\nu}^{j}(x)\right)\right>_m ,
\label{eqnmon}
\end{equation}
where $D_{\mu\nu}$ is unity on any  simple surface of plaquettes  whose 
edge is the Wilson loop, and $<\;>_{m}$ means the average over the
ensemble of magnetic currents. 
The dual field strength is defined by
${\raise0.20ex\hbox{*}}F_{m\mu\nu}^{j}
=\epsilon_{\mu\nu\alpha\beta}F_{m\alpha\beta}^{j}/2$.  Before extraction of
potentials, the $<W_{mon}^{j}>$ are averaged over color.

\pagestyle{plain}
\section{Maximal Center Gauge}
\label{sec_center}

Gauge transformations involving a discrete subgroup are 
well-defined when acting on a path ordered
exponential of gauge fields,
$$
U(x_{1},x_{2}) \equiv P \left(\exp(\int_{x_{1}}^{x_{2}} A_{\mu}dx_{\mu})\right).
$$
Hence
a gauge can be sought where a set of $U(x_{1},x_{2})$
are as close as possible to elements of a discrete  subgroup of the gauge
group.

In lattice gauge
theory, the $U(x_{1},x_{2})$ are taken to be the links of the lattice.
The direct maximal center gauge for $SU(N)$ lattice gauge theory
seeks maxima of the functional
\be
{\bf G}_{dmcg}^{suN}=\sum_{x,\mu} |{\rm tr}U_{\mu}|^{2},
\label{Ndcg}
\ee
over the $SU(N)$ gauge group.
Since ${\rm tr}U_{\mu}$ is real for $SU(2)$,
an alternate form for $SU(2)$ is
\be
{\bf G}_{dmcg}^{su2}= \sum_{x,\mu} \left( {\rm tr}U_{\mu}\right)^{2}.
\ee
The $U_{\mu}$ are in the fundamental 
representation in all the formulae above. 
Fundamental and adjoint traces are related by 
\mbox{$ {\rm tr}(U_{\mu}^{adj})=|{\rm tr} U_{\mu}|^{2} -1$,}  so Eq.(\ref{Ndcg}) can be rewritten as
\be
{\bf G}_{dmcg}^{suN}=\sum_{x,\mu}\left( {\rm tr} (U_{\mu}^{adj})+1\right),
\ee
which shows  that the DMCG can be thought of as an adjoint  
 Landau gauge. 
Gauge transformations are made so as to make 
the gauge-fields as small as possible, but  any $Z(N)$ factors in the
fundamental links are left unfixed.

Although we will restrict ourselves to the bilinear condition
Eq.(\ref{Ndcg}) in this work, it is worth noting that
for $N>2$, there are additional 
$Z(N)$ invariant functionals of the links that can 
be constructed \cite{greensite}.  
For example in 
$SU(3)$ , the expressions ${\rm tr}(U_{\mu}U_{\mu}U_{\mu})$, 
${\rm tr}(U_{\mu}U_{\mu})\cdot {\rm tr}U_{\mu}$, and 
${\rm tr}(U_{\mu})\cdot {\rm tr}(U_{\mu}) \cdot {\rm tr}(U_{\mu})$ 
are all $Z(3)$ invariant.   From these a family of real-valued, 
$Z(3)$ invariant functionals
of the links could be written down and used to define $SU(3)$ maximal 
center gauges.
It would be interesting to know if such gauges differ in their predictions
from Eq.(\ref{Ndcg}).

\subsection{Indirect Maximal Center Gauge}
In this paper, we take the indirect route to the maximal center gauge
for $SU(3)$.
This is the same method first used for $Z(2)$ center projection 
in $SU(2)$ \cite{greensite2}, and fits naturally with our study of the MAG
for $SU(3)$. A desirable feature of
the IMCG is that it allows a study of
the relationship between monopoles and center vortices.
We start by going  to the MAG as discussed in
Sec.(\ref{sec_mag}), using the functional of  Eq.(\ref{lattmag}).   
The 
resulting $SU(3)$  links  $U_{\mu}$ are then projected to 
$U(1) \times U(1)$ links $u_{\mu}$ using the methods
discussed in Sec.(\ref{abel_proj}).  Parameterizing the $u_{\mu}$ as in
Eq.(\ref{su3diag}), and using them to evaluate  Eq.(\ref{Ndcg}), 
the functional to be maximized for the IMCG becomes
\begin{eqnarray}
{\bf G}_{imcg}^{su3} & =  {\displaystyle \sum_{x,\mu}} & \left[ 3 +
2\cos(\phi^{11}_{\mu}(x)-\phi^{22}_{\mu}(x))
+2\cos(\phi^{11}_{\mu}(x)+2\phi^{22}_{\mu}(x))\right.
\label{icgcond} \\
& &\left. +2\cos(\phi^{22}_{\mu}(x)+2\phi^{11}_{\mu}(x)) \right], 
\nonumber
\end{eqnarray}
where the unitarity condition 
has been used to eliminate $\phi_{\mu}^{33}$.
The gauge functional  of Eq.(\ref{icgcond})
is maximized by successively applying abelian 
gauge transformations  at each site.
The  gauge transformations are parameterized 
as 
$$g(x)=e^{i\tilde{\lambda}_{8}\alpha_{8}(x)} e^{i\lambda_{3}\alpha_{3}(x)},$$
where
$$\tilde{\lambda_{8}} \equiv {\rm diag}(1,1,-2), \;
\lambda_{3} \equiv {\rm diag}(1,-1,0).$$
Maximizing  
at a site  $x$ gives two coupled equations 
in $\alpha_{3}(x)$ and $\alpha_{8}(x)$. These equations are not algebraically
soluble, and their solution is found iteratively.
Making the required gauge transformation at $x$  generally causes 
the maximization conditions at the nearest neighbors of $x$ to be violated, so
reaching a local maximum of ${\bf G}_{imcg}^{su3}$
requires repeated cycles through the lattice.
At a maximum,   two gauge conditions are satisfied.
The first comes from the
$\lambda_{8}$ term in the $U(1) \times U(1)$ gauge transformation,
\begin{equation}
\sum_{\mu}  \partial^{-}_{\mu} \left[\sin(2\phi^{11}_{\mu}(x) 
+\phi^{22}_{\mu}(x))
+\sin(2\phi^{22}_{\mu}(x) + \phi^{11}_{\mu}(x))\right] = 0.
\label{icgcond1}
\end{equation}
The second comes from the $\lambda_{3}$ term,
\begin{equation}
\sum_{\mu} \partial^{-}_{\mu} 
 [2\sin(\phi^{11}_{\mu}(x)-\phi^{22}_{\mu}(x)) 
 +\sin(2\phi^{11}_{\mu}(x)+\phi^{22}_{\mu}(x))
 -\sin(2\phi^{22}_{\mu}(x)+\phi^{11}_{\mu}(x))] = 0.
 \label{icgcond2}
\end{equation}
In the naive continuum limit these two conditions 
reduce
to  the abelian  Landau gauge conditions
$$\partial_{\mu} A_{\mu}^{8}=0, \;
\partial_{\mu} A_{\mu}^{3}=0,$$
where $aA_{\mu}^{8}=(\phi^{11}_{\mu}(x)+\phi^{22}_{\mu}(x))/2$,
and $aA_{\mu}^{3}=(\phi^{11}_{\mu}(x)-\phi^{22}_{\mu}(x))/2$.
In our lattice calculations, the right hand sides of Eqs.(\ref{icgcond1})
and (\ref{icgcond2}) are not actually zero, but are very small.  
This is discussed in Sec.(\ref{sec_z3results}).

After a local maximum is reached to sufficient accuracy, 
the  links $u_{\mu}$ are projected to the nearest element of
the center group, $Z(3)$, where
$$Z(3) \in \{ \exp(-i 2\pi/3), 1, \exp(+i2\pi/3)   \}\times {\bf 1}.$$ 
This results in a set of $Z(3)$-valued links, $z_{\mu}(x)$.  These are used
to compute the $Z(3)$ Wilson loops discussed in Sec.(\ref{sec_z3results}).

Since our results for center projection are all for the IMCG,
it is worth commenting on possible differences between the direct and
indirect maximal center gauges.
The IMCG depends on the specific functional form used for the MAG functional, 
which as discussed in Sec.(\ref{sec_mag}), is not unique for $SU(3)$.   
This means  the IMCG  results obtained in this paper 
using the Eq.(\ref{lattmag}) form of the MAG are
not necessarily indicative of those that would be obtained in the DMCG for \sth.
For the case of an $SU(2)$ gauge group, where the 
MAG functional is  unambiguous, we  have found
no statistically significant difference between indirect and direct
methods of reaching the maximal center gauge\cite{stacktucker}.  
A similar result may hold in the end for $SU(3)$, but this will
probably require an MAG functional different from Eq.(\ref{lattmag}).

\pagestyle{plain}
\section{$SU(3)$ Results}
\label{full_su3}
Our calculations were performed  at couplings $\beta=5.9$ and 6.0
on lattices of size  \mbox{ $10^{3}\times 16$} and $16^{4}$, respectively.
The correlation length as determined by the string tension is
approximately $4a$ at $\beta=5.90$ and $4.5a$ at $\beta=6.0$, so
at $\beta=5.90$, our $10^{3}\times 16$ 
lattice is over twice the correlation length
in size, while at $\beta=6.0$, the $16^4$ lattice
is more than three times as large as the correlation  length.  Finite
size effects should be small, particularly at $\beta=6.0$.
In each case,  $SU(3)$ 
configurations were generated using the Wilson action with a heatbath 
update and overrelaxation \cite{kennpenn}.  The first several thousand 
configurations were discarded to allow for equilibration, and 
every $20^{\rm th}$ configuration after that was used for measurements. 
We analyzed 440 configurations at $\beta=5.9$, and 600 at $\beta=6.0$.

In this section, we present full $SU(3)$ results for the heavy quark 
potential.  Results 
for the MAG $U(1)\times U(1)$ , MAG monopole 
and IMCG $Z(3)$ potentials are presented in the sections 
immediately following.
The full \sth potential is of course gauge-invariant, and therefore not
subject to the uncertainties of gauge-fixing and projection.
On the other hand, the noise level of \sth Wilson loops is much
greater than for projected loops, and variance reduction is essential to
extract the potential.  The presence of a large Coulomb term also
complicates the determination of the string tension.
For the \sth potential, modified Wilson loops were computed using a
multihit variance
reduction scheme \cite{parisi} and spatial smearing \cite{ape}.  
In computing  the heavy quark potential, we used loops 
of spatial extent 
$R=2a$ to $R=5a\,(\beta=5.9)$,  $R=2a$ to $R=7a\,(\beta=6.0)$.
The maximum temporal extent was $T=12a$ in both cases.
From the Wilson loops, we extracted
potentials by fitting $-\ln(W(R,T))$ to a straight line in
$T$, the slope giving $V(R)$ \cite{js_su2}.  
The values of $T$ included in the fits 
for each $R$ 
were $T=R+a$ to $T=12a$.  The potentials were then fitted to
the form $V(R) = V_{0}+\alpha /R+\sigma R$.    

Our results for the full \sth potential parameters are presented 
in Table \ref{full_results}.
For the purposes of this paper, the string tensions 
in Table \ref{full_results}  are the most significant,
since the various projection schemes we will compare to in later sections 
do not (nor are they expected to)
reproduce the Coulomb part of the potential.  
Comparing our \sth results to
the modern high statistics calculations available 
in the literature, our $\beta=6.0$ string
tension is within error bars of the values quoted in 
\cite{michaelpot,balipot1,balipot2,balipot3}.  At $\beta=5.90$, the only
calculation with comparable accuracy is \cite{born}, where the
value reported is somewhat 
higher than ours, by an amount slightly outside of error bars.
However, it should be noted that at $\beta=6.0$, the 
string tension quoted in \cite{born}
is also somewhat higher than the results 
in \cite{michaelpot,balipot1,balipot2,balipot3}.  We conclude that our
full \sth string tensions are of good quality, comparable to the best values
available in the literature.

\section{ $U(1) \times U(1)$ Results}
\label{sec_u1u1}
This section contains our most important results.   Abelian projection
after gauge-fixing to the MAG retains the largest number of degrees of
freedom of the various projections we consider, and so is a less severe
truncation of the theory than
retaining only the abelian fields
of monopoles, or keeping only $Z(3)$-valued links after center projection.

The process of actually finding a local maximum of the MAG functional was 
carried out
iteratively, determining the gauge transformation $g(x)$ at a specific site
which maximizes the terms in ${\bf G}_{mag}^{su3}$ 
that are affected by $g(x)$.  Gauge
transformations were successively chosen from the three 
independent subgroups of $SU(3)$.  An arbitrary $SU(3)$  gauge transformation
can be expressed
as a product of such $SU(2)$ gauge transformations.
To improve convergence, we used an
overrelaxation procedure, replacing $g(x)$ with $g^{\omega}(x)$ 
\cite{mandula}.  For gauge transformations which are elements of
$SU(2)$
subgroups, it is possible to work out $g^{\omega}(x)$ exactly, rather 
than use the Taylor expansion approximation described in \cite{mandula}.
It was found that 
$\omega=1.85$ was the optimum overrelaxation parameter when using the
exact expression for $g^{\omega}(x)$.  
Referring back to the conditions discussed after Eq.(\ref{lattmag}), 
gauge 
fixing iterations were performed until 
\be
\langle |{\bf X}(x)_{12}|^{2}+ |{\bf X}(x)_{21}|^{2}
+|{\bf X}^{\prime}(x)_{13}|^{2}+ |{\bf X}^{\prime}(x)_{31}|^{2}
+|{\bf X}^{\prime\prime}(x)_{23}|^{2}+ |{\bf X}^{\prime\prime}(x)_{32}|^{2} 
\rangle \,\leq 10^{-12},
\label{gaugecond}
\ee
where $< >$ implies a lattice average.
Eq.(\ref{gaugecond}) is
a very tight constraint. We are confident that when it is
satisfied, the configuration is very close to a
local maximum of  ${\bf G}_{mag}^{su3}$.  As with any maximal gauge,
the functional ${\bf G}_{mag}^{su3}$ has many local maxima or gauge copies.
A single gauge-fixing brings a configuration near a random local maximum
of the functional, unrelated in general to the global maximum.  Actually
finding the global maximum is impractical.  However, it is  feasible
to generate a sample of gauge copies for each configuration, and
study how the projected string tension varies when 
copies with steadily increasing values of the functional are used. 
Our results are presented first for the
case of one gauge copy/configuration, followed by analysis of the
effect of gauge ambiguities.

After reaching a local maximum  for each
configuration, the $SU(3)$ links were
projected to  the $U(1)\times U(1)$ links $u_{\mu}$, using the methods
discussed in Sec.(\ref{abel_proj}).  
Spatial links were smeared 
using the Ape method \cite{ape} with a staple weighting parameter of 
$0.7$, iterating the procedure 8 times per link. 
Then abelian Wilson loops were computed using
\be
W_{abel}(R,T)=\left\langle{1 \over 3}\left({\rm tr}\prod_{ C}u_{\mu}(x)\right)\right\rangle.
\label{u1wilson}
\ee
These loops were 
used to compute the $U(1) \times U(1)$  heavy quark potential $V(R)$ 
at both $\beta=5.90$ and $\beta=6.0$.  The results for fits to $V(R)$
of the same type as for the full $SU(3)$ potential are shown in
Table \ref{u1u1_results}.

The $U(1)\times
U(1)$ string tensions shown 
in Table \ref{u1u1_results}  are for one 
local maximum or gauge copy/configuration, and
are  smaller than the
corresponding full \sth results, by 10\% 
at $\beta=6.0$.  This is in
contrast to the behavior of $SU(2)$ in the MAG for one gauge copy/configuration.
There similar gauge-fixing methods
produce  $U(1)$ string tensions at least 10\%  larger than full 
$SU(2)$ at $\beta_{SU(2)}=2.4,2.5,2.6$ \cite{bornyakovsu2,bali}.  

Turning to gauge ambiguities, 
in order for their effect to bring the $U(1) \times
U(1)$ string tension closer to the full \sth answer, local maxima of
the gauge functional with higher values of the functional would have
to produce {\it higher} abelian projected string tensions.  This is the 
opposite of what happens in $SU(2)$. 
To explore whether this unexpected behavior occurs  for the $SU(3)$ MAG
functional  of Eq.(\ref{lattmag}),  we
generated  $N_{copy}=10$
gauge copies/configuration for each of 200 configurations at $\beta=6.0$.  
To generate a gauge copy, a move is made to an arbitrary  point 
on the gauge orbit of
a configuration by making a random gauge transformation.  The gauge-fixing
algorithm then takes the configuration to a nearby local maximum.  
Due to abelian projection, the $U(1)\times U(1)$ string tension is dependent
on which copies are used for
each configuration.  Our treatment of the variation  due to gauge copies
follows the method  of Bali, et. al. \cite{bali}, developed for the
$SU(2)$ case.
For each configuration, 
a subset of size $n_{g}$ of the $N_{copy}$ 
copies is chosen.  
Only the configuration in the subset with the highest value of
the gauge functional is retained and used to calculate Wilson loops.  
The Wilson loops are averaged over
all such subsets of size $n_{g}$. 
For $n_{g}=1$,  this is just the
average  over all $N_{copy}$ 
copies for each configuration, whereas for $n_{g} = N_{copy}$,
only the copy with the highest value of the functional for all $N_{copy}$ 
copies is used for each configuration.  
The resulting Wilson loops are then averaged
over configurations  and used to calculate potentials.
The output of this analysis is a set of 
values of the $U(1) \times U(1)$ string tension
for $n_{g}$ ranging from 1 to $N_{copy}$.  
The results are shown
in Table \ref{u1_gribov_results}.  As can be seen there, 
the $U(1) \times U(1)$ string tension steadily decreases as $n_{g}$ increases.
At $n_{g}=N_{copy}=10$,  the $U(1) \times U(1)$
string tension has decreased by
 10\%, and the ratio of the $U(1) \times U(1)$ to the $SU(3)$ string tension is
0.80(4).

We have thus found that
the effect of gauge ambiguities  in $SU(3)$ is similar to
$SU(2)$, namely, the higher the value of the MAG functional,
the lower the abelian projected string tension.  
In $SU(2)$, this trend is beneficial,
since for a randomly chosen gauge copy, the $U(1)$ string tension 
is above the full \stw result, and taking gauge copies with successively 
higher values of the functional brings the string tension down, closer and
closer to the full $SU(2)$ value.
For $SU(3)$,
using the Eq.(\ref{lattmag}) form of the MAG, 
the $U(1) \times U(1)$ string tension from a random gauge copy 
is too low to start with, and 
gauge copies with higher functional values produce $U(1) \times U(1)$
string tensions still further
below the $SU(3)$ string tension.

\pagestyle{plain}

\section{Monopole Results}
\label{sec_monoresults}

Before presenting results on the heavy quark potential due to monopoles, we
discuss a  point mentioned in Sec.(\ref{mono_loc}), related to
the determination of the magnetic current.
Our procedure for locating 
monopoles produces a separate current $m_{\mu}^{j}$ for each color,
$j=1,2,3$.  This method treats the three colors symmetrically,
and is not prejudiced toward a specific group structure for the
monopoles.  Only two of these currents are really independent.
Although it does not vanish 
link by link,  
the total current, $m_{\mu}^{1}+m_{\mu}^{2}+m_{\mu}^{3}$,
is in effect a null current. 
We have checked this by using the total current 
to calculate monopole Wilson loops.  The resulting string 
tension (at $\beta=5.90$) is zero to four decimal places,
$\sigma_{mon}^{tot}=0.0000(4)$, as it should be.

Turning to the monopole potentials, at $\beta=5.9$ and $6.0$, 
we computed monopole Wilson loops for
each color using Eq.(\ref{eqnmon}), averaged the results over color,
and extracted heavy quark 
potentials using the same fit methods discussed earlier.  
The results are
presented in Table \ref{u1u1_mono_results}  along with the current density,
or fraction $f_{m}$ of 
links carrying magnetic current.   The results shown in 
Table \ref{u1u1_mono_results}  are for one MAG gauge-fixing/configuration, 
and are on the same footing as the $U(1)\times U(1)$ results in 
\mbox{Table \ref{u1u1_results}.}  The inequality \mbox{$\sigma_{SU(3)} > 
\sigma_{U(1)\times U(1)} > \sigma_{mono}$} is seen to hold 
for both $\beta$ values,  the monopole string tensions
being approximately $ 25\%$ smaller than the $SU(3)$ string tensions,
a discrepancy well outside of error bars.  

It is of interest to compare these results with corresponding
ones for $SU(2)$.  The full $SU(2)$ string tensions at $\beta_{SU(2)}=2.4$ and 
$\beta_{SU(2)}=2.45$ are numerically close 
to the full $SU(3)$ string tensions at
$\beta_{SU(3)}=5.9$ and $\beta_{SU(3)}=6.0$, respectively, implying
similar correlation lengths.  However, 
for $SU(2)$, the
monopole string tensions for one gauge-fixing/configuration are within
$5\%$ of the full $SU(2)$ values \cite{stackwensleyneiman}.  Further, the
magnetic current densities $f_{m}$, are more than twice as large as the 
corresponding  $SU(3)$ results shown in Table \ref{u1u1_mono_results} .
Thus for $SU(3)$, the MAG condition of Eq.(\ref{lattmag}) leads to projected 
configurations with
a severe shortage of monopoles.
Despite this shortage, the magnetic current
density shows scaling properties rather
similar to those found in $SU(2)$.  If the magnetic current density is a
physical density, dividing it by the cube of a physical energy should 
produce a scaling quantity, i.e. one independent of $\beta$.  
Computing  $f_{m}/(\sigma_{mono})^{3/2}$ for $\beta=5.90$ and $\beta=6.0$ gives
1.1(1) and 1.0(1), respectively.
Corresponding results for $SU(2)$
at $\beta_{SU(2}=2.40$ and $\beta_{SU(2}=2.45$ are 1.6(1) and 1.7(1),
respectively \cite{stackwensleyneiman}.  
So although the magnetic current density for $SU(3)$ is
significantly smaller, it scales with $\beta$ in a similar way to the
$SU(2)$ case when the monopole string tension defines the unit of energy.

Just as they affect the $U(1) \times U(1)$ potential, MAG 
gauge ambiguities affect the monopole  potential parameters.  Using the
same set of $N_{copy}=10$ gauge copies/configuration at $\beta=6.0$, 
we can determine how the monopole string tension and magnetic current
density vary with the number of gauge copies.  The analysis proceeds
exactly as in the $U(1) \times U(1)$ case.  The results are presented in
Table \ref{mono_gribov_results}.  As $n_{g}$ increases, the monopole
string tension decreases by 
essentially  the same factor as 
the $U(1)\times U(1)$ string tension, and the
ratio of the magnetic current density to 
$(\sigma_{mon})^{3/2}$ remains approximately constant.
By the time $n_{g}=N_{copy}=10$ has been reached, 
the $\beta=6.0$ value of the monopole string tension
 is 34\% smaller than the $SU(3)$ string tension.

\section{$Z(3)$  Results}
\label{sec_z3results}

To transform an $SU(3)$ configuration to the 
indirect maximal center gauge requires two gauge-fixings and one
projection.  First a local maximum of the MAG functional is found, followed
by abelian projection.  The resulting  $U(1) \times U(1)$
$u_{\mu}$  links are then gauge-fixed by finding a local maximum
of the IMCG functional Eq.(\ref{icgcond}), using the
procedure described in Sec.(3.1).  Iterations were 
performed until the sum of the 
squares of Eqs.(\ref{icgcond1}) and 
(\ref{icgcond2}) was less than $10^{-14}$, when averaged over 
the lattice.  During the gauge fixing we monitored the modulus of the 
abelian trace, $|{\rm tr} u_{\mu}|$.  If all 
links were elements of the center, the value of ${1 \over 3}\langle|{\rm 
tr}u_{\mu}|\rangle$ would be unity.
At $\beta=5.90$,  we observed 
it to change from an average value of 0.525 before gauge fixing 
to a value of 0.922 after gauge 
fixing.  The behavior at $\beta=6.0$ is similar.

The IMCG has gauge ambiguities of two types; those inherent in the MAG, as 
well as those that are associated with the second gauge-fixing just described.
We first present results for one gauge-fixing of each type/configuration.  This
puts each configuration in a random local maximum of both gauge conditions.
The resulting center-projected  links 
were used to compute $Z(3)$ Wilson loops
and extract a heavy quark potential as in the previous cases.  Similar to
the monopole case, the noise level is very low, and the potential is
almost perfectly linear.  Table \ref{icg_results}  shows the
potential parameters, along with $p$, the fraction of plaquettes with
non-trivial $Z(3)$ flux.  
These plaquettes are
conjectured to represent the geometrical
centers of finite-sized center vortices. 
Usually called P-vortices, we
simply refer to  non-trivial $Z(3)$ plaquettes  
as vortices.  
From Table \ref{icg_results}  and Table \ref{full_results},
the ratio of the $Z(3)$ string tension to the 
full \sth value is $ 0.88(8)$ at $\beta=5.9$, but falls to $0.80(8)$ at
$\beta=6.0$.  

The vortex density is an areal density, so if it is physical, dividing
by the square of a physical energy should produce a scaling quantity.
The natural quantity to divide by is the $Z(3)$ string tension.
Computing the ratio of the vortex density to the $Z(3)$ string tension  
gives .73(1) and .75(1) for $\beta=5.90$ and $\beta=6.0$ respectively.
For the DMCG in $SU(2)$ the results at $\beta_{SU(2}=2.40$ and 
$\beta_{SU(2}=2.50$ are quite similar; 0.76(1) and 0.78(1), respectively
\cite{stacktucker}.  Thus for both $SU(3)$ and $SU(2)$, dividing 
the vortex density by the 
$Z(N)$ string tension removes most of the $\beta$-dependence and produces
an approximate scaling quantity.

We also explored the effect of gauge ambiguities on
the IMCG results.  To avoid an expensive simultaneous 
treatment of both MAG and IMCG gauge ambiguities, 
we generated 10 IMCG  
gauge copies only for the MAG copy with the highest value of the MAG
functional.   This is the best MAG copy from the viewpoint of the MAG
functional.  
The analysis of the $Z(3)$ string tension proceeded 
in exactly the same manner as
the $U(1) \times U(1)$  case.  
The results are shown in
Table \ref{ICG_gribov_results}, where it is seen that the $Z(3)$ string
tension decreases by O(10\%) in going from $n_{g}=2$ to $n_{g}=10$.  The
ratio of vortex density to $Z(3)$ string tension is approximately constant.
At $n_{g}=10$, 
value of the $Z(3)$ string tension is roughly 40\% smaller than
 the $\beta=6.0$  $SU(3)$ string tension.

If a different choice of MAG copy had been made, it is reasonable to 
expect a similar O(10\%) variation due to IMCG gauge ambiguities.  
If for example, a random MAG gauge copy is used for each configuration and
IMCG copies generated, the $Z(3)$ string tension would be expected to 
decrease from its value given in Table \ref{icg_results}  
to one O(10\%) lower, so for
$n_{g}=10$, the $Z(3)$ string tension would be O(70\%) of
the $\beta=6.0$ $SU(3)$ string tension.

A reasonable conclusion is that using the gauge 
functionals Eq.(\ref{lattmag}) for the MAG
and Eq.(\ref{icgcond}) for the IMCG, 
whatever choice is made of MAG or IMCG gauge copies, the result is a
$\beta=6.0$ $Z(3)$  string tension whose central value is at
least 20\% smaller than the full $SU(3)$ string tension.

Finally, it is of interest to study the connection
between vortices and monopoles.  Suppose the magnetic current  is
non-zero on a dual lattice link.  This link is dual to a
3-cube on the original lattice.
The partial magnetic current density $f_{m}(n)$ is defined 
as the current density for
links where the dual cube has $n$ faces pierced by $Z(3)$ vortices. 
The total magnetic current density is then
$$
f_{m}=\sum_{n=0}^{6}f_{m}(n).
$$
Since $Z(3)$ center vortices form closed surfaces, 
$f_{m}(1)\equiv 0$, but all other possibilities can
and do occur.  
The study was carried out for one gauge-fixing of each
type/configuration.  We computed the ratio $r_{m}(n)=f_{m}(n)/f_{m}$, which 
measures the fractional occurrence for the cube dual to a magnetic current link
to have $n$ faces pierced by $Z(3)$ vortices.  The results are shown in
Table \ref{mono_vort_results}, and hold independent of color.
Thus approximately 85\% (83\%) of the time at $\beta=5.9\,(6.0)$, 
the cube dual to  a magnetic current link has  at least two faces
pierced by vortices.  
This says there is an intimate connection between center vortices and
monopoles in \sth, similar to what was found previously in \stw
\cite{stacktucker, greensitemono}.   It could in fact be true that  the
actual fractions are 100\%, with the
smaller numerical results caused by inaccuracies in locating monopoles and
vortices on the lattice.

\pagestyle{plain}
\section{Summary and Conclusions}
\label{sec_conc_js}
\begin{figure}[hbt]
\includegraphics[width=0.80\textwidth]{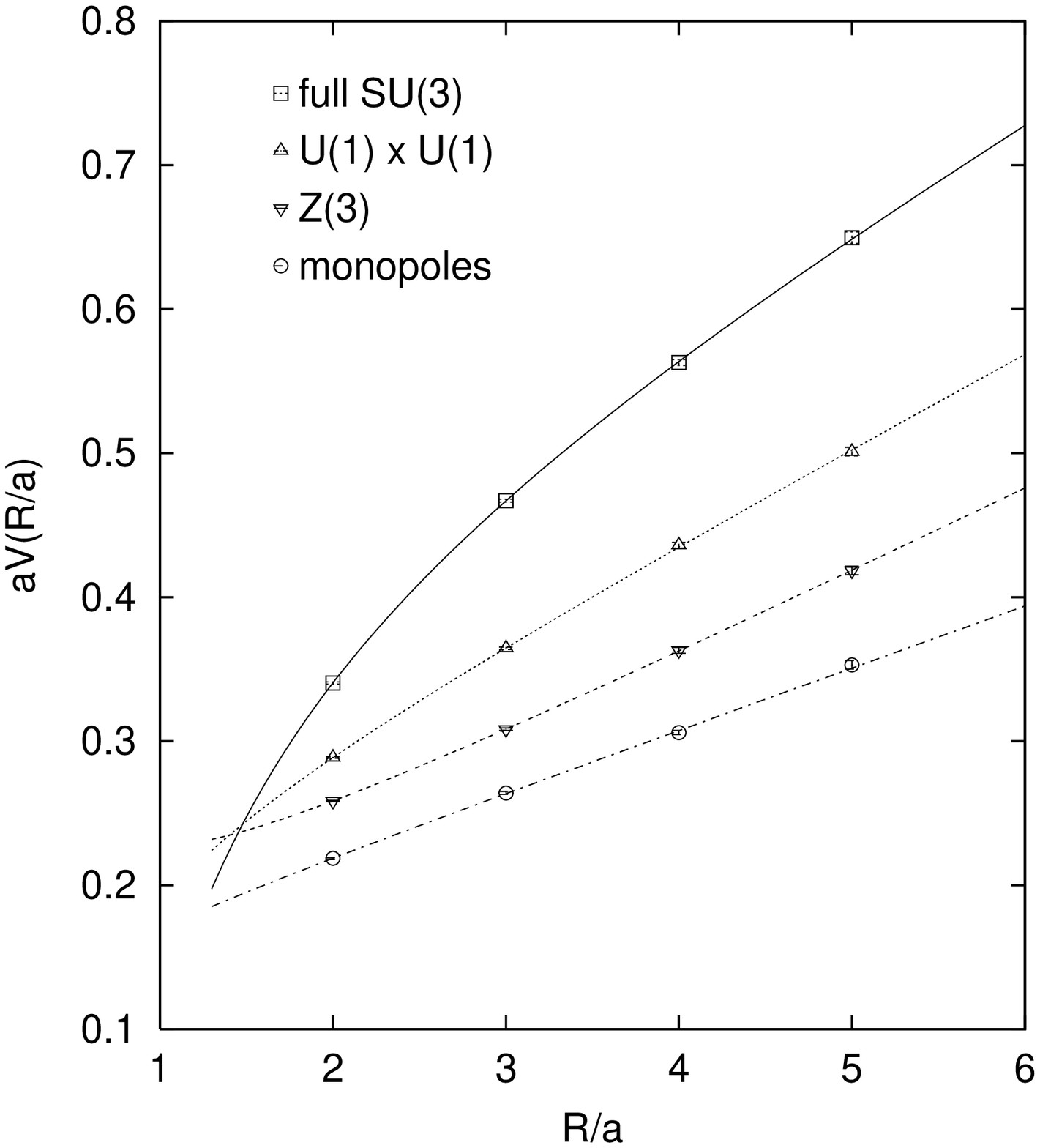}
\caption{Heavy quark potentials at $\beta=5.9$ for full $SU(3)$, 
$U(1)\times U(1)$, $Z(3)$, and monopoles }
\label{pots59}
\end{figure}

\begin{figure}[hbt]
\includegraphics[width=0.80\textwidth]{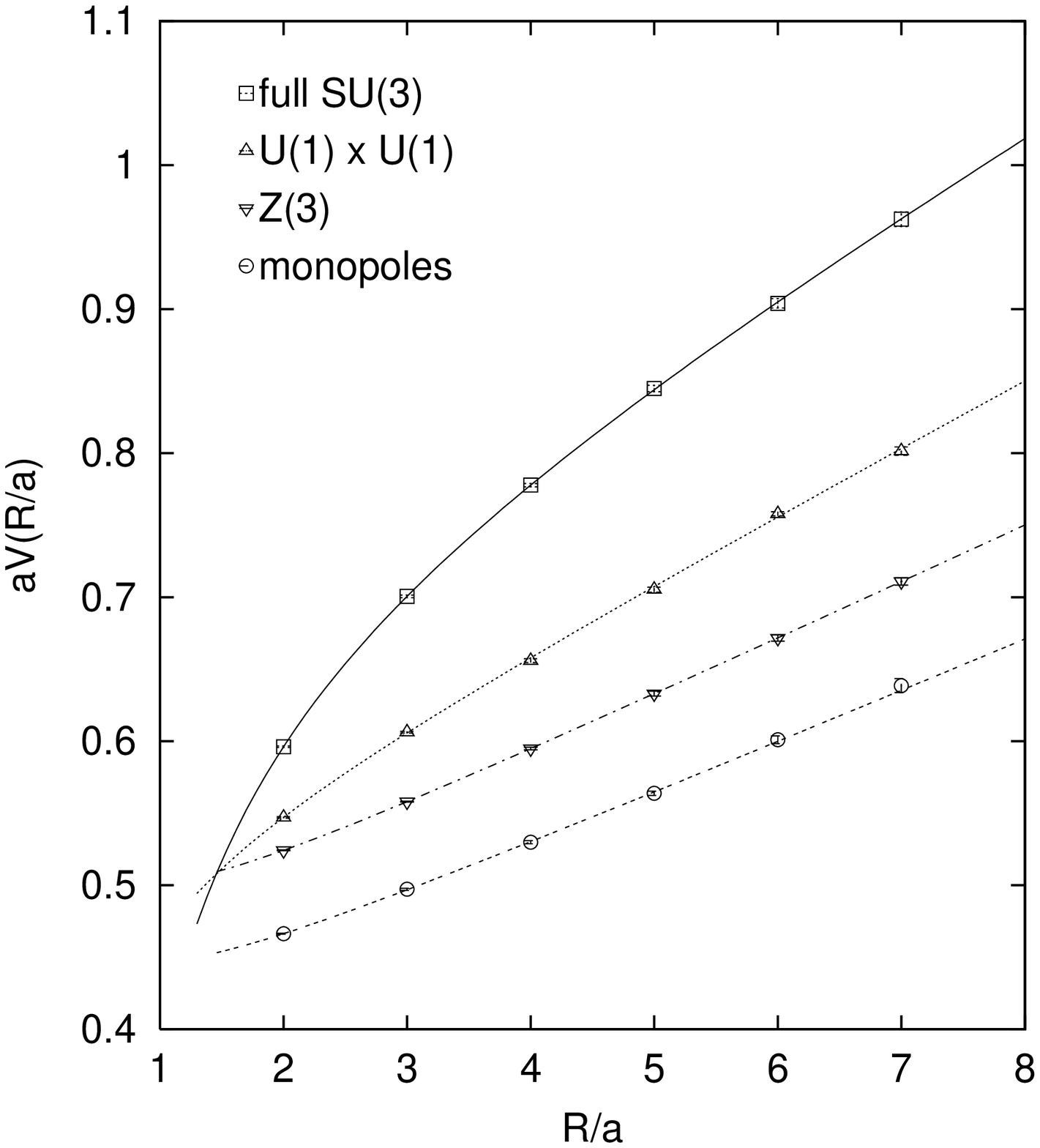}
\caption{Heavy quark potentials at $\beta=6.0$ for full $SU(3)$, 
$U(1)\times U(1)$, $Z(3)$, and monopoles }
\label{pots60}
\end{figure}

We have calculated the heavy quark potential for \sth lattice gauge
theory at \mbox{$\beta=5.90$} and $\beta=6.0$. These \sth results have been
compared with three cases which project onto a reduced set of 
degrees of freedom after gauge-fixing. 
The $U(1)\times U(1)$ and monopole results involve the 
version of the MAG expressed in Eq.(\ref{lattmag}), 
while the $Z(3)$ results involve
in addition
the IMCG gauge-fixing of Eq.(\ref{icgcond}).  In Figs.(\ref{pots59}) 
and (\ref{pots60}),
we show plots of the full $SU(3)$ and projected potentials, for the
case of one gauge-fixing/configuration.  The potentials have been shifted
by constants for convenient display, the significant quantity to compare being
the slope of each potential at large R, which determines the string tension.
The  string tension results obey the inequalities
\mbox{ $\sigma_{SU(3)} > \sigma_{U(1)\times U(1)} 
> \sigma_{Z(3)} > \sigma_{mono}$.}
These inequalities hold at one gauge-fixing/configuration  and become
stronger when account is taken of gauge ambiguities.  The most important
result of this paper is that 
\mbox{$\sigma_{SU(3)} > \sigma_{U(1)\times U(1)}$.}
When account is taken of gauge ambiguities at $\beta=6.0$, 
the $U(1) \times U(1)$ string tension is smaller than the full $SU(3)$
string tension by 20\%, an amount well outside of 
error estimates.  If  a simulated annealing gauge-fixing
algorithm \cite{bornyakovsu2,bali} were used 
instead of the overrelaxation method used here, the
discrepancy could be still larger.
This situation for $SU(3)$  
is in contrast to $SU(2)$, where  the MAG and
extrapolating to the global maximum of the functional ($n_{g} \rightarrow
\infty$) produces
a $U(1)$ string tension very close to the full \stw result \cite{bornyakovsu2}.

It is conceivable that for $SU(3)$, the discrepancy between $\sigma_{SU(3)}$ and
$\sigma_{U(1) \times U(1)}$ could fade away with increasing $\beta$ and
become negligible in the continuum limit.  This seems unlikely given that
the ratio $\sigma_{U(1) \times U(1)} / \sigma_{SU(3)}$ is smaller at
$\beta=6.0$ than $\beta=5.9$.   Rather than moving into the region $\beta>6.0$
using the Eq.(\ref{lattmag}) form of the MAG, 
we intend to explore the possibility that 
the remedy lies in a generalized MAG such as Eq.(\ref{latthiggs}).  
The condition of 
Eq.(\ref{latthiggs}) seems to 
express in a natural way the idea that there is an
effective adjoint Higgs field present, whose direction in the Cartan algebra
may vary from point to point on the lattice.  If it is accepted that 
the simple MAG form of Eq.(\ref{lattmag}) 
cannot work for $SU(3)$, it is important to
explore more general forms of the MAG.  Failure to find any form of MAG
which works would have serious implications for `t Hooft's  proposal of
abelian dominance of confinement 
for $SU(N)$, $N>2$.  On the other hand, 
a quantitatively successful generalized MAG in \sth opens the possibility of
a similar result for the indirect maximal center gauge, and a detailed 
study of the contribution to confinement of MAG monopoles and IMCG
center vortices.  These two degrees of freedom are clearly intimately
connected.

With any generalized MAG, there will be a rather clear signal for improvement.
An improved form of the MAG should
produce a string tension {\it larger} than the full  \sth result for
one gauge-fixing/configuration. 
Gauge copies with higher values of the functional will inevitably 
lower the abelian
$U(1) \times U(1)$ string tension,
hopefully to within error bars of the full \sth result.
In contrast,
a form of the MAG which for one gauge-fixing/configuration 
produces an abelian projected string
tension  {\it smaller} than the full \sth result is  untenable. 
Gauge copies with higher values of the functional will 
only reduce it further.  Our claim is that the  MAG in the form of
Eq.(\ref{lattmag}) is of this type.

With regard to center projection, since  
$Z(3)$ is a subgroup
of $U(1)\times U(1)$,   
the group center should be accessible indirectly, i.e. after abelian 
projection.  However, 
given the subtleties of the MAG for $SU(3)$, a thorough study of the
direct maximal center gauge for $SU(3)$ is quite desirable.  
The DMCG avoids the question of what is the proper form of the MAG.  
We intend to study the direct maximal center gauge for 
$SU(3)$ in future work.

\subsection*{Acknowledgements}
This work was supported in part by the National Science
Foundation under grants NSF PHY 9900658 (JDS) and NSF PHY 9802579 (RJW).  
The numerical computations were supported in part by NSF cooperative
agreement ACI-9619020 through computing resources provided by the
National Partnership for Advanced Computational Infrastructure at
the San Diego Supercomputer Center and the  Texas Advanced Computing Center.



\begin{table}[htb]
\caption{Full \sth results for string tension, Coulomb coefficient,
and self energy at $\beta=5.90$ and $\beta=6.0$ }
\vspace{0.5cm}
\begin{center}
\begin{tabular}{||c|c|c||}\hline
$SU(3)$  & $\beta=5.90$ & $\beta=6.0$ \\ \hline
Vol & $10^{3}\times 16$ & $16^4$ \\ \hline
$\sigma$ & 0.068(3) & 0.050(1) \\ \hline
$\alpha$  & -0.36(2) & -0.32(2) \\ \hline
$V_{0}$  &  0.68(2)     & 0.66(2)  \\ \hline
R/a        & 2 to 5   & 2 to 7   \\ \hline
\end{tabular}
\end{center}
\label{full_results}
\end{table}

\begin{table}[htb]
\caption{MAG $U(1) \times U(1)$ results for string tension, Coulomb coefficient,
and self energy at $\beta=5.90$ and $\beta=6.0$, one gauge copy/config. }
\vspace{0.5cm}
\begin{center}
\begin{tabular}{||c|c|c||}\hline
$U(1)\times U(1)$ & $\beta=5.90$ & $\beta=6.0$ \\ \hline
Vol & $10^{3}\times 16$ & $16^4$ \\ \hline
$\sigma$ & 0.063(3) & 0.045(2) \\ \hline
$\alpha$  & -0.09(2) & -0.07(2) \\ \hline
$V_{0}$  &  0.20(2)     & 0.19(2)  \\ \hline
R/a        & 2 to 5   & 2 to 7   \\ \hline
\end{tabular}
\end{center}
\label{u1u1_results}
\end{table}

\begin{table}[htb]
\caption{$\beta=6.0$ MAG $U(1)\times U(1)$ string tensions for subsets of
size $n_{g}$ out of 10 gauge copies}
\vspace{0.5cm}
\begin{center}
\begin{tabular}{||c|c||} \hline
$n_{g}$ & $\sigma_{U(1)\times U(1)}$ \\ \hline
  2        & 0.042(1) \\ \hline
  4        & 0.041(1) \\ \hline
  6        & 0.041(1) \\ \hline
  8        & 0.040(1) \\ \hline
  10        & 0.040(1) \\ \hline
\end{tabular} 
\end{center}
\label{u1_gribov_results}
\end{table}

\begin{table}[htb]
\caption{MAG monopole results for string tension, Coulomb coefficient,
self energy, and current density,  at $\beta=5.90$ and $\beta=6.0$,
one gauge copy/configuration}
\vspace{0.5cm}
\begin{center}
\begin{tabular}{||c|c|c||}\hline
 $mono$& $\beta=5.90$ & $\beta=6.0$ \\ \hline
Vol & $10^{3}\times 16$ & $16^4$ \\ \hline
$\sigma$ & 0.050(2) & 0.038(1) \\ \hline
$\alpha$  &  0.03(2) &  0.045(1) \\ \hline
$V_{0}$  &  -0.04(2)     & -0.051(4)  \\ \hline
$f_{m}$      & $1.28(1) \times 10^{-2}$ & $7.45(3) \times 10^{-3}$ \\ \hline
R/a        & 2 to 6   & 2 to 8   \\ \hline
\end{tabular}
\end{center}
\label{u1u1_mono_results}
\end{table}

\begin{table}[htb]
\caption{$\beta=6.0\;$ MAG monopole string tensions and current densities 
for subsets of
size $n_{g}$ out of 10 gauge copies}
\vspace{0.5cm}
\begin{center}
\begin{tabular}{||c|c|c||} \hline
$n_{g}$ & $\sigma_{mono}$ & $f_{m}$ \\ \hline
  2        & 0.036(1) & $7.25(4)\times 10^{-3}$ \\ \hline
  4        & 0.035(1) & $7.08(4)\times 10^{-3}$ \\ \hline
  6        & 0.034(1) & $7.00(4)\times 10^{-3}$ \\ \hline
  8        & 0.034(1) & $6.95(4)\times 10^{-3}$ \\ \hline
  10        & 0.033(1)&  $6.92(5)\times 10^{-3}$ \\ \hline
\end{tabular} 
\end{center}
\label{mono_gribov_results}
\end{table}

\begin{table}[htb]
\caption{IMCG results for $Z(3)$ string tension, Coulomb coefficient,
self energy, and vortex density,  at $\beta=5.90$ and $\beta=6.0$,
one gauge copy/configuration }
\vspace{0.5cm}
\begin{center}
\begin{tabular}{||c|c|c||}\hline
$IMCG$ & $\beta=5.9$ & $\beta=6.0$ \\ \hline
Vol & $10^{3}\times 16$ & $16^4$ \\ \hline
$\sigma$ & 0.060(3) & 0.040(2) \\ \hline
$\alpha$  &  0.05(2) &  0.03(1) \\ \hline
$V_{0}$  &  -0.04(2)     & -0.02(1)  \\ \hline
$p$      & $4.40(1) \times 10^{-2}$ & $3.01(1) \times 10^{-2}$ \\ \hline
R/a        & 2 to 6   & 2 to 8   \\ \hline
\end{tabular}
\end{center}
\label{icg_results}
\end{table}

\begin{table}[htb]
\caption{$\beta=6.0$ IMCG $Z(3)$  string tensions and vortex densities 
for subsets of
size $n_{g}$ out of 10 IMCG gauge copies, best MAG gauge copy}
\vspace{0.5cm}
\begin{center}
\begin{tabular}{||c|c|c||} \hline
$n_{g}$ & $\sigma_{Z(3)}$ & $p$ \\ \hline
  2        & 0.033(1) & $2.73(4)\times 10^{-2}$ \\ \hline
  4        & 0.030(1) & $2.67(4)\times 10^{-2}$ \\ \hline
  6        & 0.029(1) & $2.64(4)\times 10^{-2}$ \\ \hline
  8        & 0.029(1) & $2.62(4)\times 10^{-2}$ \\ \hline
  10        & 0.029(1)&  $2.61(4)\times 10^{-2}$ \\ \hline
\end{tabular} 
\end{center}
\label{ICG_gribov_results}
\end{table}

\begin{table}[htb]
\caption{ For cube dual to magnetic current link, $r_{m}(n)\equiv$
fraction of time $n$ faces are pierced by $Z(3)$ flux}
\vspace{0.5cm}
\begin{center}
\begin{tabular}{||c|c|c||} \hline
  & $\beta=5.90$ & $\beta=6.0$ \\ \hline
  $r_{m}(0)$        &  0.14(1) & 0.16(1) \\ \hline
  $r_{m}(2)$        &  0.74(1) & 0.74(1) \\ \hline
  $r_{m}(3)$        &   0.09(2) & 0.07(2) \\ \hline
  $r_{m}(4)$        &   0.02(2) &  0.02(2) \\ \hline
  $r_{m}(5)$        &  0.002(2)&  0.001(1) \\ \hline
  $r_{m}(6)$        & 0.001(1)&  0.001(1)\\ \hline
\end{tabular} 
\end{center}
\label{mono_vort_results}
\end{table}


\end{document}